\documentclass[useAMS,usenatbib]{mn2e}

\usepackage{graphicx}
\usepackage{times}
\usepackage{natbib}
\usepackage{amsmath}
\usepackage{hyperref}

\newcommand{\lesssim}{\la}

\title[Early structures and critical metallicity]{The transition from population III to population II-I star formation}
\author[Umberto Maio et al.]{
Umberto~Maio$^{1,2}$\thanks{E-mail:
umaio@mpe.mpg.de, maio@mpa-garching.mpg.de (UM)}
Benedetta~Ciardi$^1$,
Klaus~Dolag$^1$,
Luca~Tornatore$^3$,
Sadegh~ Khochfar$^2$\\
${}^1$Max-Planck-Institut f\"ur Astrophysik,
Karl-Schwarzschild-Stra{\ss}e 1, D-85748 Garching b. M\"unchen, Germany\\
${^2}$
Max-Planck-Institut f\"ur extraterrestrische Physik,
Giessenbachstra{\ss}e 1, D-85748 Garching bei M\"unchen, Germany\\
${^3}$
Dipartimento di Astronomia dell'Universit\'a di Trieste,
via Tiepolo 11, I-34131 Trieste, Italy
}

\begin{document}

\date{...(draft)}
\pagerange{\pageref{firstpage}--\pageref{lastpage}}\pubyear{}
\maketitle
\label{firstpage}


\begin{abstract}

We present results from the first cosmological simulations which study the
onset of primordial, metal-free (population III), cosmic star formation
and the transition to the present-day, metal-rich star formation  
(population II-~I), including molecular (H$_2$, HD, etc.) evolution, 
tracing the injection of metals by supernov{\ae} into the
surrounding intergalactic medium and following the change in the
initial stellar mass function (IMF) according to the metallicity of the
corresponding stellar population. 
Our investigation addresses the role of a wide variety of parameters
(critical metallicity for the transition, IMF slope and range, SN/pair-instability 
SN metal yields, star formation threshold, resolution, etc.) on the metal-enrichment 
history and the associated transition in the star formation mode.
All simulations present common trends. Metal enrichment is very patchy,
with rare, unpolluted regions surviving at all redshifts, inducing the   
simultaneous presence of metal-free and metal-rich star formation regimes.
As a result of the rapid pollution within high-density regions due
to the first SN/pair-instability SN, local metallicity is quickly
boosted above the critical metallicity for the transition.
The population III regime lasts for a very short period during the first stages of star formation  ($\sim 10^7\,\rm yr$), and its average contribution to the total star formation rate density drops rapidly below $\sim 10^{-3}-10^{-2}$.

\end{abstract}




\section{Introduction}\label{sect:intro}

The standard paradigm of cosmic structure formation relies on the
classical approach of Jeans' theory \cite[]{Jeans1902} applied to
primordial matter fluctuations in the frame of an expanding Universe.
Cosmological models for structure formation have been developed since
several decades \cite[e.g.][]{GunnGott1972,Peebles1974,WhiteRees1978}
and the overall picture agrees with a ``flat'' Universe where
``cold-dark matter'' is the dominant fraction of matter and the
``cosmological constant'', $\Lambda$, is the dominant fraction of the
cosmological energy density. Baryonic structures arise from in-fall
and condensation of gas into dark-matter potential wells. In
particular, it seems \cite[e.g. ][]{Maiolino_et_al_2007} that
molecular gas could account for a significant fraction of the
dynamical mass of early objects. Recent determinations of the
cosmological parameters \cite[]{wmap7_2010} suggest a present-day
expansion rate $H_0\simeq 70\rm~km/s/Mpc$ (in units of $\rm
100~km/s/Mpc$ this parameter becomes $h\simeq 0.70$), a total-matter
density parameter $\Omega_{0m}\simeq 0.272$, with ``baryonic''
component $\Omega_{0b}\simeq 0.0456$, and a cosmological-constant
density parameter $\Omega_{0\Lambda}\simeq 0.728$. The primordial
power spectrum of perturbation is well fitted by a power law with
index $n\simeq 0.96$ and normalization via mass variance within
$8$~Mpc/$h$ radius $\sigma_8\simeq 0.8$. As a reference, it is common
to define the standard $\Lambda$CDM model the one with the following
parameters:
$H_0=\rm 70~km/s/Mpc$,
$\Omega_{0m}=0.3$, $\Omega_{0b}=0.04$,
$\Omega_{0\Lambda}=0.7$,
$\Omega_{0tot}=1.0$,
$n=1$,
$\sigma_8=0.9$.
In the frame of cosmic evolution, it is believed that structure
formation takes place from the growth of primordial fluctuations in
matter density. These would contract and collapse allowing gas cooling
and the subsequent build-up of stars. The determination of the
properties of early stars and their effects on the following
baryonic-structure formation episodes is still a problem under debate
\cite[for a complete review see e.g.][]{Ciardi_Ferrara_2005}.
Nevertheless, it is reasonably well established that the very first
generation of stars should be characterized by massive objects with
typical masses much larger than the presently observed ones
\cite[][]{SS1953, WW1995, Larson1998, Chiosi2000,
  HegerWoosley2002,HegerWoosley2008}. These primordial stars
(population III stars) are formed out of a pristine environment, where
the cooling agents are limited to primordial, H-based molecules only,
i.e. H$_2$ and HD, which are able to cool the gas down to temperatures
of $\sim 10^2~\rm K$. Therefore, the mass of primordial stars should
be relatively large and their spectrum, biased towards such large
objects, is commonly referred to as ``top-heavy'' initial mass
function (IMF) \cite[][]{Larson1998}. These features imply very short
lifetimes (up to $\sim 10^6$ years only) and final death mostly into
black holes \cite[][]{HegerWoosley2002}. The only mass range where
primordial stars can explode as pair-instability supernov{\ae} (PISN)
and pollute the surrounding medium is [140-260]~M$_\odot$
\cite[][]{HegerWoosley2002}. Nevertheless, also at masses of $\sim
100-140~\rm M_\odot$ there can be some mass loss before collapse
(pulsational pair SN). The key uncertainty here is primary nitrogen
production and the dredge up of carbon and oxygen. In particular, if
the stellar atmosphere is highly CNO enhanced there may be substantial
mass loss, but zero metallicity should still be a good first
approximation for such stellar flows (S.~Woosley, private
communication).
\\ 
Despite the many uncertainties on their
characteristics, population III (popIII) stars have an important
impact on the evolution of the intergalactic medium (IGM), since they
initiate the metal pollution of the IGM, with consequent change of its
chemical composition and cooling properties (chemical feedback).
Therefore, star formation events in enriched regions will happen in
completely different conditions, because metals allow further cooling
and fragmentation to smaller scales. This results in an initial
stellar-mass function peaked at lower masses, similar to the nowadays
observed Salpeter-like IMF \cite[]{Salpeter1955} for population II-I
(popII-I) stars.\\ A very debated issue is the understanding of the
transition from the primordial popIII star formation regime to the
standard popII-I regime. In this respect, there are evidences for the
existence of a critical metallicity, $Z_{crit}$, at which the
modalities of star formation allow such transition
\cite[][]{Bromm_et_al_2001,Schneider_et_al_2002}: $Z_{crit}$ is the
metallicity at which the metal cooling function dominates over the
molecular one. In this case, the fragmentation process becomes highly
enhanced, but the exact value is not well-established, yet. Different
studies suggest discrepant values with $Z_{crit}$ varying between
$\sim 10^{-6}~Z_\odot$ \cite[e.g.][]{Schneider_et_al_2006} and $\sim
10^{-3}~Z_\odot$ \cite[e.g.][]{Bromm_Loeb_2003}\footnote{
We adopt $Z_\odot\simeq 0.0201$
\cite[][]{Anders_Grevesse_1989, Grevesse_Sauval_1998}. See, however,
\cite{Asplund_et_al_2009} for a recently updated value of $Z_\odot\simeq 0.0134$}.
\\
The main uncertainty is the presence of dust at high redshift.
To our knowledge, dust is produced and injected in the ISM mainly
 by the low--mass stars in the AGB phase.
 This would imply no dust production at very-high redshift, due to 
``long'' stellar lifetimes: a 8~M$_\odot$ star has a life of $\sim 0.1$ Gyr,
comparable with the age of Universe at $z\sim 15-20$. None the less,
the presence of large amounts of dust and heavy elements has been detected 
at moderately-high redshift, i.e. at $z>6$, when the Universe is younger than 
$\sim 1~\rm Gyr$ \cite[e.g. ][]{Bertoldi2003,Maiolino_et_al_2004}.
This suggests that dust production must have occurred primarily in the ejecta of
supernova explosions, which are the final fate of massive, short-lived
stars. If, indeed, supernov{\ae} could be able to induce dust
production then star formation in the early Universe and the level of
$Z_{crit}$ would be strongly influenced, as well. Enrichment by dust
would not impact the thermal structure of the IGM, but its presence
would alter the whole star formation process. Indeed, when the
metallicity of star forming regions is still below $\sim
10^{-6}\,Z_\odot$ the only relevant coolants are molecules, mainly
H$_2$ and HD, while above $\sim 10^{-4}\,Z_\odot - 10^{-3}\,Z_\odot$
gas cooling is fully dominated by metal fine-structure transitions
\cite[e.g. ][]{Maio2007} and cloud fragmentation can happen down to
sub-solar scales
\cite[][]{Bromm_et_al_2001,Schneider_et_al_2002,SFS2004}. In between,
cooling capabilities depends mainly on the amount of metals depleted
onto dust grains. The mass content locked into dust grains, almost
independently from the exact mass of SNII/PISN progenitor, reaches
$2\%-5\%$ of the parent mass for type-II SN
\cite[][]{Kozasa1991,TodiniFerrara2001,Nozawa_et_al_2003,BianchiSchneider2007},
and $15\%-30\%$ for PISN \cite[][]{Nozawa_et_al_2003,SFS2004}, in a
few hundred days after the beginning of the explosion. Despite the
huge difference between the dust mass fractions deriving from SNII and
PISN, the dust-to-metal mass ratios (or ``depletion factors'') turn
out to be $\sim 0.3 - 0.7$ in both cases and, for metal poor SNII
progenitors with masses of $\rm 25\,M_\odot - 30\, M_\odot$, it is up
to $\sim 1$.
The natural conclusion is that first massive stars could spread in the early
Universe a lot of dust during their final phases (as either SNII or PISN).
\\
In the present work, we aim at investigating the birth of the first
stars, the following cosmic metal enrichment from their explosive
death and the transition to the standard, presently observed, star
formation regime. We focus on how such transition and the associated
features are affected by different choices of $Z_{crit}$, IMF, post-supernova metal yields,
 star formation density thresholds, box dimension and resolution.
\\
Throughout the paper we will refer to
``population III regime'' and ``population II-I regime'' when the
metallicity is below or above the critical level, respectively. In
addition, we will adopt the standard $\Lambda$CDM cosmological
model.\\ The paper is organized as follows: we describe the code and
the simulations in Sect. \ref{sect:sims}, results are given is
Sect. \ref{sect:results}, and parameter dependence is addressed in
Sect. \ref{sect:params} (\ref{sect:imf}, \ref{sect:threshold}, and
\ref{sect:largerboxes}). We discuss and conclude in Sect. \ref{sect:disc}.


\section{Simulations}\label{sect:sims}

\begin{table*}
\centering
\caption[Simulation parameters]{Parameters adopted for the simulations.}
\begin{tabular}{lccccccccc}
\hline
\hline
Model & box side & number of & mean inter-particle & $M_{gas}\rm [M_\odot/{\it h}]$ & $M_{dm}\rm [M_\odot/{\it h}]$ & $Z_{crit}$ & $n_{\rm H, th}$ & popIII IMF& Massive\\
& [Mpc/{\it h}] & particles & separation [kpc/$h$] & & & &  $\rm [cm^{-3}]$ & range [M$_\odot$]& yields\\
\hline
Zcrit3 & 0.7 &$2\times 320^3$ & 2.187 & $1.16\times 10^2$ & $7.55\times 10^2 $ & $10^{-3}Z_\odot$   & 70 & [100, 500]& HW02$^1$\\
Zcrit4 & 0.7 &$2\times 320^3$ & 2.187 & $1.16\times 10^2$ & $7.55\times 10^2 $ & $10^{-4}Z_\odot$   & 70 & [100, 500]& HW02$^1$\\
Zcrit5 & 0.7 &$2\times 320^3$ & 2.187 & $1.16\times 10^2$ & $7.55\times 10^2 $ & $10^{-5}Z_\odot$   & 70 & [100, 500]& HW02$^1$\\
Zcrit6 & 0.7 &$2\times 320^3$ & 2.187 & $1.16\times 10^2$ & $7.55\times 10^2 $ & $10^{-6}Z_\odot$   & 70 & [100, 500]& HW02$^1$\\
\hline
Zcrit4-r8-100 & 0.7 &$2\times 320^3$ & 2.187 & $1.16\times 10^2$ & $7.55\times 10^2 $ & $10^{-4}Z_\odot$   & 70 & [8, 100]& HW08$^2$\\
Zcrit4-r0.1-100 & 0.7 &$2\times 320^3$ & 2.187 & $1.16\times 10^2$ & $7.55\times 10^2 $ & $10^{-4}Z_\odot$   & 70 & [0.1, 100]& HW08$^2$\\
\hline
Zcrit4-HW8-40 & 0.7 &$2\times 320^3$ & 2.187 & $1.16\times 10^2$ & $7.55\times 10^2 $ & $10^{-4}Z_\odot$   & 70 & [0.1, 100]& HW08$^2$\\
Zcrit4-WW8-40 & 0.7 &$2\times 320^3$ & 2.187 & $1.16\times 10^2$ & $7.55\times 10^2 $ & $10^{-4}Z_\odot$   & 70 & [0.1, 100]& WW95$^3$\\
\hline
Zcrit4-th7 & 0.7 &$2\times 320^3$ & 2.187 & $1.16\times 10^2$ & $7.55\times 10^2 $ & $10^{-4}Z_\odot$   & 7 & [100, 500]& HW02$^1$\\
Zcrit4-th1 & 0.7 &$2\times 320^3$ & 2.187 & $1.16\times 10^2$ & $7.55\times 10^2 $ & $10^{-4}Z_\odot$   & 1 & [100, 500]& HW02$^1$\\
Zcrit4-th01 & 0.7 &$2\times 320^3$ & 2.187 & $1.16\times 10^2$ & $7.55\times 10^2 $ & $10^{-4}Z_\odot$   & 0.1 & [100, 500]& HW02$^1$\\
\hline
Zcrit4-5th7 & 5.0 &$2\times 320^3$ & 15.62 & $4.23\times 10^4$ & $ 2.75\times 10^5$ & $ 10^{-4}Z_\odot$  & 7 & [100, 500]& HW02$^1$\\
Zcrit4-10th7 & 10.0 &$2\times 320^3$ & 31.25 & $3.39\times 10^5$ & $2.20\times 10^6$ & $10^{-4}Z_\odot$  & 7 & [100, 500]& HW02$^1$\\
\hline
\hline
\label{tab:sims}
\end{tabular}
\begin{flushleft}
{\small
$^1$ Data tables from \cite{HegerWoosley2002}.\\
$^2$ Data tables from \cite{HegerWoosley2008}.\\
$^3$ Data tables from \cite{WW1995}.
}
\end{flushleft}
\end{table*}
The simulations were performed in the frame of the standard
$\Lambda$CDM cosmological model, by using the Gadget-2 code
\cite[][]{Springel2005}. The initial conditions for the reference
simulations were generated sampling 0.7~Mpc/$h$, comoving (but see
also discussion on larger boxes in Sect. \ref{sect:largerboxes}) of
the cosmic fluid at redshift $z=100$, with $320^3$ particles per gas
and dark matter species, having masses of $116\,{\rm M_\odot}/h$ and
$755\,{\rm M_\odot}/h$, respectively.
We perform the numerical simulations including primordial molecular evolution
\cite[][]{Maio2006,Maio2007,MaioPhD}, star formation prescriptions
\cite[][]{MaioPhD,Maio2009}, and metal pollution from stars
\cite[][]{TBDM2007}. We choose the stellar IMF according to the value
of the gas metallicity $Z$: if $Z<Z_{crit}$ a top-heavy IMF is
assumed, otherwise a Salpeter one.\\ More in detail, we include in the
code the whole set of chemistry reactions leading to molecule creation
or destruction and metal pollution, with the corresponding cooling
functions. We follow the abundances of the H- and He-based chemistry,
involving e$^-$, H, H$^+$, H$^-$, He, He$^+$, He$^{++}$;
H$_2$-chemistry following H$_2$, H$_2^+$ \cite[][]{Yoshida_et_al_2003}
and deuterium chemistry with D, D$^+$, HD, and HeH$^+$
\cite[][]{Maio2007}. While in principle an arbitrary number of metals
can be accounted for by the scheme for metallicity tracking
\cite[][]{TBDM2007}, in this work we focus on C, O, Mg, S, Si, and Fe,
since they are the most abundant metals produced during stellar
evolution. We assume, as initial chemical composition, a primordial
neutral gas with residual electron and H$^+$ fractions \cite[][]{GP98}
$x_{\rm e^-}\simeq x_{\rm H^{+}}\simeq 4\cdot 10^{-4}$,
H$_2$ fraction $x_{\rm H_2}=10^{-6}$,
H$_2^+$ fraction $x_{\rm H_2^+}=3\cdot 10^{-21}$,
D fraction $x_{\rm D}=3.5\cdot 10^{-5}$,
D$^+$ fraction $x_{\rm D^+}=4\cdot 10^{-9}$,
HD fraction $x_{\rm HD}= 7\cdot 10^{-10}$,
HeH$^+$ fraction $x_{\rm HeH^+}=   10^{-14}$.
Individual metals are set to zero initial abundance.\\
The natural gas evolution, with the initial in-fall into the
dark-matter potential wells and the shock-heating phase, is fully
self-consistently followed through the attainment of the isothermal
turn-over and the subsequent cooling regime.
This stage is very important at early times, as it represents the state when gas cooling balances and takes over heating.
Resolving the isothermal phase in primordial-chemistry simulations is fundamental to account for the early molecular cooling of the gas \cite[as already discussed and shown in][]{Maio2009}.
For this reason, we assume that stars are formed roughly at the end of the
gas cooling process, when a suitable physical density threshold for
early star formation of $\sim 10^2\rm~cm^{-3}$ (more exactly $70\,\rm
cm^{-3}$) is reached \cite[][]{MaioPhD,Maio2009}. The often used
low-density threshold of $\sim 0.1 \rm~cm^{-3}$
\cite[e.g.][]{Katz1996} would induce spurious time anticipation of the
onset of star formation \cite[][]{Maio2009}.
Hereafter, stars evolve and die according to their characteristic lifetimes,
possibly expelling newly created metals into the surrounding gas and
modifying its cooling properties either at low \cite[e.g.][]{Maio2007} and at high temperatures \cite[e.g.][]{SutherlandDopita1993,Wiersma2009, Cantalupo2010}.
\\
For the very first bursts (and those happening in gas of metallicity smaller than
$Z_{crit}$) we adopt a ``Salpeter-like'' IMF, in the mass range
$[100,\,500]\,\rm M_\odot$, with a slope of $-1.35$. The stellar
life-times lie between $\sim 3\cdot 10^6\,\rm yr$, for the $100\,\rm
M_\odot$ stars, and $\sim 0\,\rm yr$ (instantaneous death), for the
hundreds-solar-masses stars.
As mentioned in the Introduction, the relevant mass range for metal pollution 
is approximatively $[140,\,260]\,\rm M_\odot$, because of the formation of PISN.
We point out that in our approach each star forming SPH particle can be
assimilated to a simple stellar population, with coeval stars formed
out from the same environment. Once the particle (i.e. stellar
population) metallicity reaches the critical value, $Z_{crit}$,
popII-I star formation is assumed to set in, in place of popIII, and
subsequent stars will have masses in the range $[0.1,\,100]\,\rm
M_\odot$, distributed according to a Salpeter IMF. PopII-I stellar
life-times are usually much longer than those of popIII, ranging
between $\sim 2\cdot 10^{10}\,\rm yr$, for the $0.1\,\rm M_\odot$
stars, and $\sim 3\cdot10^6\,\rm yr$, for the $100\,\rm
M_\odot$.
\\
Yields of popIII stars (see discussion in Sect. \ref{sect:imf}) are assumed accordingly to
\cite{HegerWoosley2002}, while for popII-I stars we follow
\cite{WW1995} for massive stars (SNII), \cite{vdH_G_1997} for mass
loss of low- and intermediate-mass stars, and
\cite{Thielemann_et_al_2003} for SNIa.\\ In order to describe properly
stellar evolution, it is important to take into account feedback
effects which deal with those physical processes that are not resolved
because of numerical limitations. In our simulations star forming
particles are subject to wind kinetic feedback\footnote{ For a
  different metal/feedback approach, see also treatments based on
  thermal feedback \cite[e.g.][]{Cecilia_et_al_2005}. } with
velocities of $500\,\rm km/s$. These particles produce metals which
are smoothed over the neighbours according to the SPH kernel.
\\
We perform four numerical simulations considering critical metallicities
of $Z_{crit}=10^{-3}Z_\odot,10^{-4}Z_\odot,10^{-5}Z_\odot,10^{-6}Z_\odot$,
which span the full range found in the literature (see Introduction).
In Sect. \ref{sect:imf}, \ref{sect:threshold}, and
\ref{sect:largerboxes} we will also investigate how the results are
affected by a different choice for the popIII IMF, the post-supernova yields, 
the critical star formation threshold and the box dimension/resolution.
A schematic description of the simulations is given in Table \ref{tab:sims}.


\section{Results}\label{sect:results}

In this section, we present results related to the early phases of
cosmological metal enrichment and the transition from popIII to
popII-I regime. The results refer to the $0.7\,\rm Mpc /{\it h}$ side
box described in the previous section, with varying $Z_{crit}$, the
most significant parameter in the calculation.
\begin{figure*}
\begin{center}
%
\includegraphics[width=0.7\textwidth]{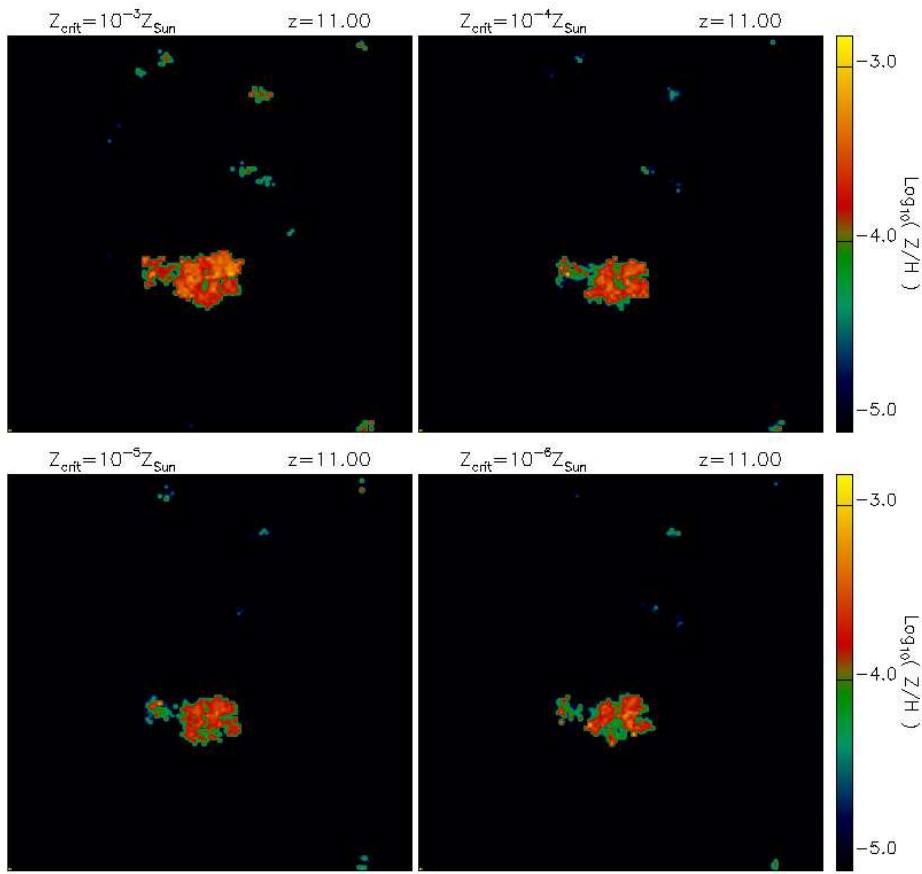}
\caption[Metallicity maps]{\small Metallicity maps at redshift $z=11$
  for the $0.7\,{\rm Mpc}/h$ side box, with $Z_{crit}=10^{-3}Z_\odot$
  (top-left panel), $10^{-4}Z_\odot$ (top-right panel),
  $10^{-5}Z_\odot$ (bottom-left panel), and $10^{-6}Z_\odot$
  (bottom-right panel). The maps are obtained by projecting the whole
  simulation box. }
\label{fig:maps_metallicity_zcrit}
\end{center}
\end{figure*}
\begin{figure*}
\begin{center}
%
\includegraphics[width=0.7\textwidth]{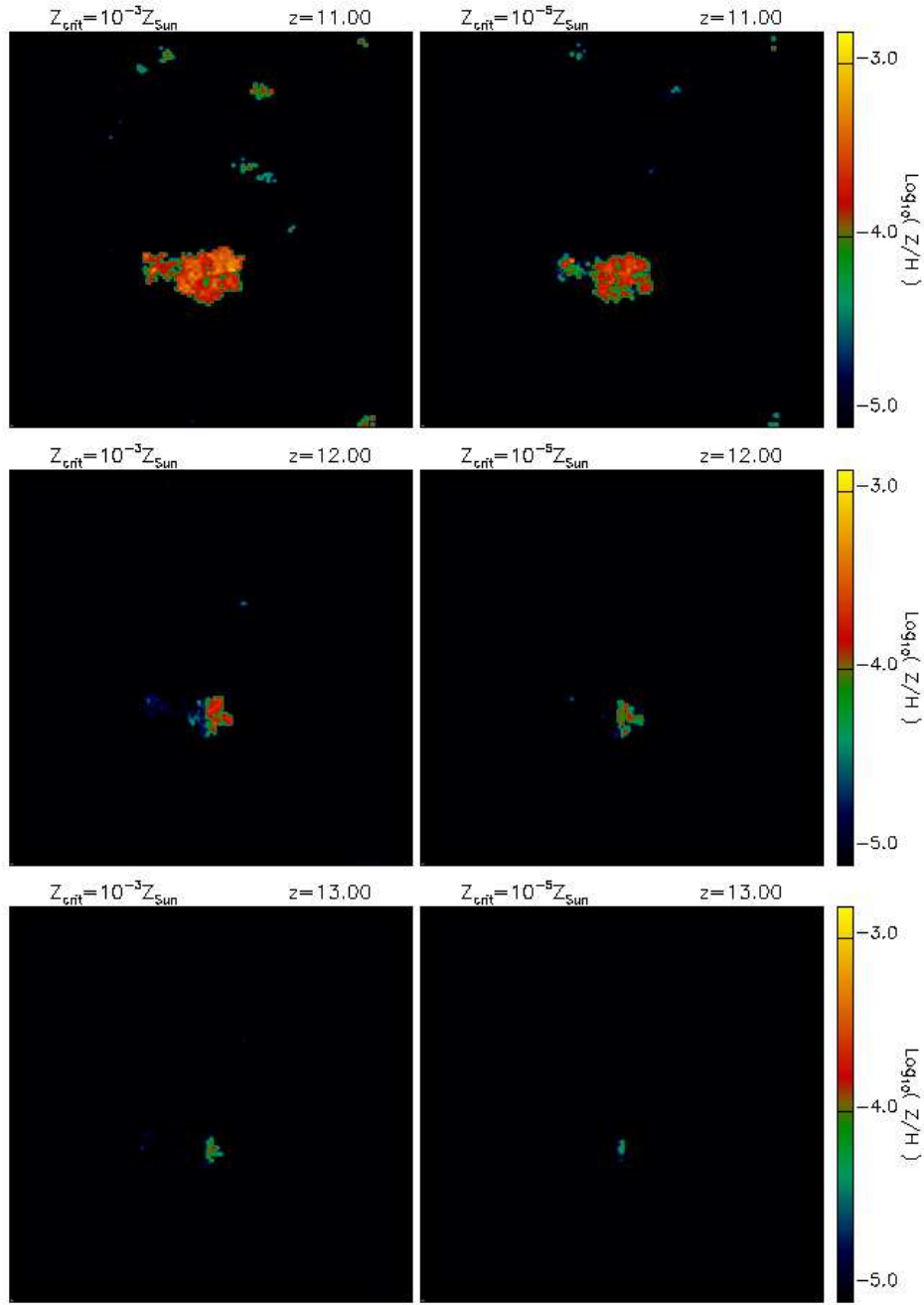}
\caption[Metallicity comparison]{\small Comparison of the metallicity
  maps at redshift $z=11$ (upper row), 12 (central row), and 13
  (bottom row), for the $0.7\,{\rm Mpc}/h$ side box with
  $Z_{crit}=10^{-3}Z_\odot$ (left column) and $10^{-5}Z_\odot$ (right
  column). The maps are obtained by projecting the whole simulation
  box. }
\label{fig:maps_metallicity_comparison}
\end{center}
\end{figure*}
\begin{figure*}
\begin{center}
%
\includegraphics[width=0.7\textwidth]{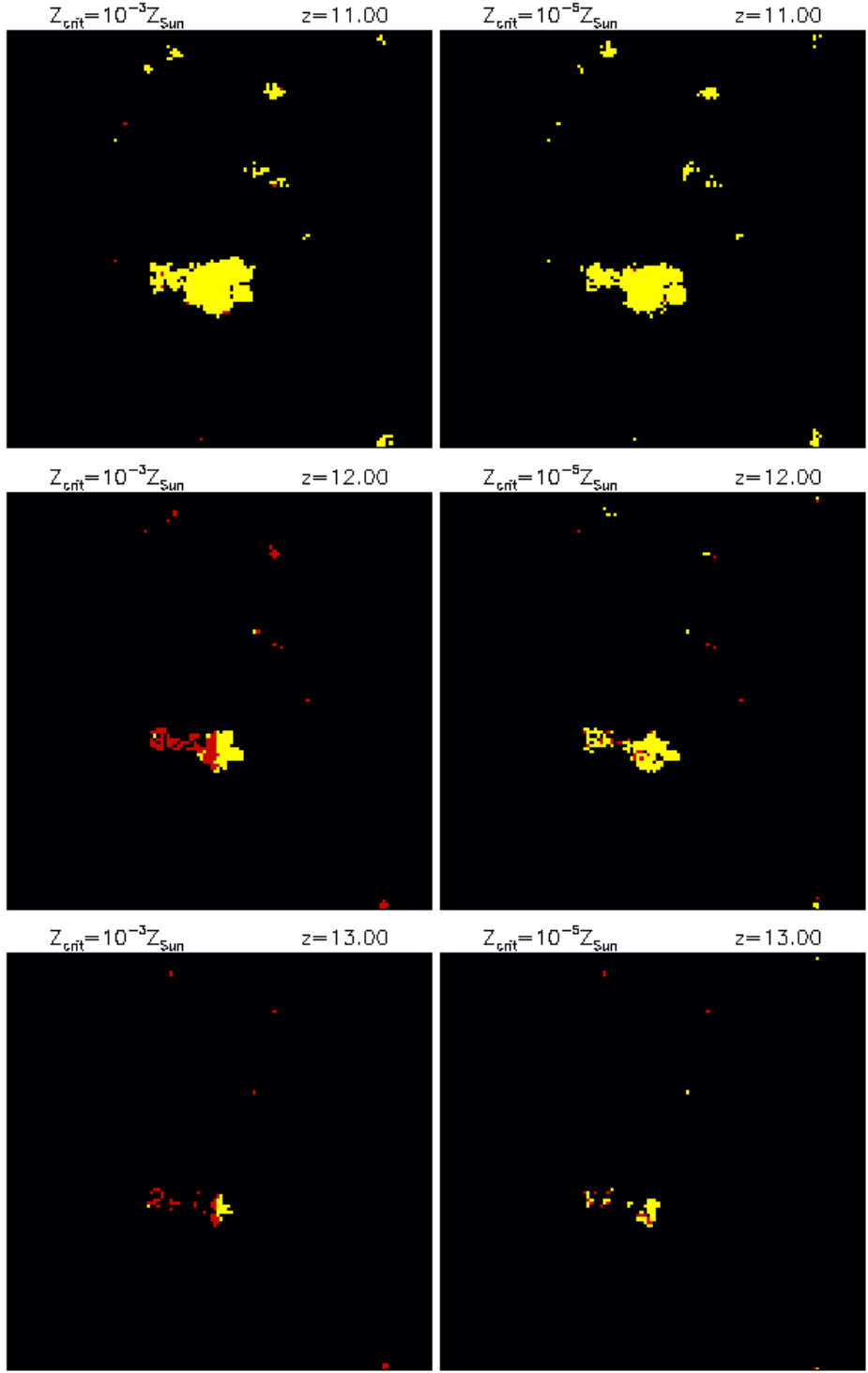}
\caption[Metallicity comparison]{\small Comparison of the
  two-population metallicity maps at redshift $z=11$ (upper row), 12
  (central row), and 13 (bottom row), for the $0.7\,{\rm Mpc}/h$ side
  box with $Z_{crit}=10^{-3}Z_\odot$ (left column) and
  $10^{-5}Z_\odot$ (right column). The yellow areas refer to polluted
  regions with metallicity higher than $Z_{crit}$, while the red ones
  to those with metallicity lower than $Z_{crit}$. The maps are
  obtained projecting the whole box. }
\label{fig:maps_metallicity_comparison_2col}
\end{center}
\end{figure*}


\subsection{Metallicity distribution}
In order to have a visual representation of the simulations we show
maps at redshift $z=11$ for the projected metallicity distribution (Fig.
\ref{fig:maps_metallicity_zcrit}) for all the cases considered:
$Z_{crit}=10^{-3}Z_\odot$ (upper-left panel),
$10^{-4}Z_\odot$(upper-right panel), $10^{-5}Z_\odot$(bottom-left
panel) and $10^{-6}Z_\odot$ (bottom-right panel). The metallicity is
defined as the metal mass fraction of each SPH particle: $Z = m_Z /
m_{tot}$, where $m_Z$ and $m_{tot}$ are respectively the mass of
metals and the total mass of each SPH particle.
\\ 
Metal enrichment is very patchy and there are regions which are very
strongly polluted and others which preserve their pristine, metal-free
composition. The main pollution events happen around the densest
regions, where star formation takes place.
We stress that the differences in the distribution and level of metallicities
are only due to the value of $Z_{crit}$ adopted. In particular, the
metallicity of the gas reaches a slightly higher value and wider
distribution for the $Z_{crit}=10^{-3}Z_\odot$ case than for the other
cases. This is due to the longer time needed to reach the critical
level, $Z_{crit}$. Once this happens, popII-I star formation sets in
and further delays metal pollution, because the SNII from popII-I stars
release about one fourth of metals per unit mass than the popIII ones,
on a timescale at least $\sim 10$ times longer.
When $Z_{crit}=10^{-6}Z_\odot$, pollution above the critical
metallicity threshold is immediate and most of the regions rapidly
move into the popII-I star formation regime. This transition gets
delayed as $Z_{crit}$ increases. \\ In Fig.
\ref{fig:maps_metallicity_comparison}, we compare metal enrichment at
different redshifts. The maps refer to $z=11$ (upper row), 12 (central
row) and 13 (bottom row), for $Z_{crit}=10^{-3}Z_\odot$ (left column)
and $10^{-5}Z_\odot$ (right column). The process is quite fast as
inferred from the rapid growth of metal-enriched bubbles, and allows
for high metal pollution within very short time (see also Sect.
\ref{sect:metallicity_evolution}).
\\
To highlight the differences between popIII and popII-I star forming regions, we plot in Fig.
\ref{fig:maps_metallicity_comparison_2col} the corresponding two-population 
maps\footnote{In this case, data are not smoothed on the edges of the pixels.} 
projecting the whole box. 
They show very well how the star formation regime evolves according to metallicity
(see further discussion in Sect. \ref{sect:SFR}).
\\ 
From a more quantitative analysis of the Figure, some distinctive behaviour emerges. For
$Z_{crit}=10^{-6}Z_\odot$, almost all the polluted areas reach the
critical metallicity already in $\sim 0.1\,\rm Gyr$ ($\sim 98\%$ at
$z=13$, $\sim 99\%$ at $z=12$, and $\sim 100\%$ at $z=11$).
In the other cases we observe a similar quick enrichment: for
$Z_{crit}=10^{-5}Z_\odot, 10^{-4}Z_\odot$ and $10^{-3}Z_\odot$ the
percentage of areas with $Z \ge Z_{crit}$ is respectively $\sim 72\%$,
$49\%$ and $32 \%$ at $z=13$; $\sim 86\%$, $ 70\%$ and $ 39\%$ at
$z=12$; and $\sim 99\%$, $97\%$ and $97\%$ at $z=11$. We note as well
that, at fixed redshift, the fraction of popII-I star formation
regions decreases with increasing $Z_{crit}$, because of the higher
enrichment level needed to allow the transition from $Z<Z_{crit}$ to
$Z\ge Z_{crit}$.\\
%
As the enrichment process is very localized, also the transition to a
different star formation regime strongly depends on location. This
means that it is possible to find coexistence of population III and
population II-I, with stronger pollution in the central regions of
star formation sites and weaker pollution in the outermost ones (see
maps in Fig. \ref{fig:maps_metallicity_zcrit} and in Fig.
\ref{fig:maps_metallicity_comparison}). It is in fact evident that the
preferential sites of popIII star formation are isolated areas and
often peripheric regions not strongly contaminated by nearby bursts.
Our findings are consistent with the study of the
$Z_{crit}=10^{-4}Z_\odot$ case by \cite{TFS2007}. We note that changes
in $Z_{crit}$ can only slightly alter the whole enrichment scenario,
as just few regions which undergo popII-I star formation for
$Z_{crit}=10^{-5}Z_\odot$ are undergoing popIII star formation for
$Z_{crit}=10^{-3}Z_\odot$ (e.g. Fig. \ref{fig:maps_metallicity_comparison_2col}).
In fact, accordingly with the previous discussion, the star forming environment 
is chemically ``sterilized'' by one or a few subsequent popIII generations.


\subsection{Metallicity evolution}\label{sect:metallicity_evolution}
\begin{figure*}
\centering
\includegraphics[width=0.4\textwidth]{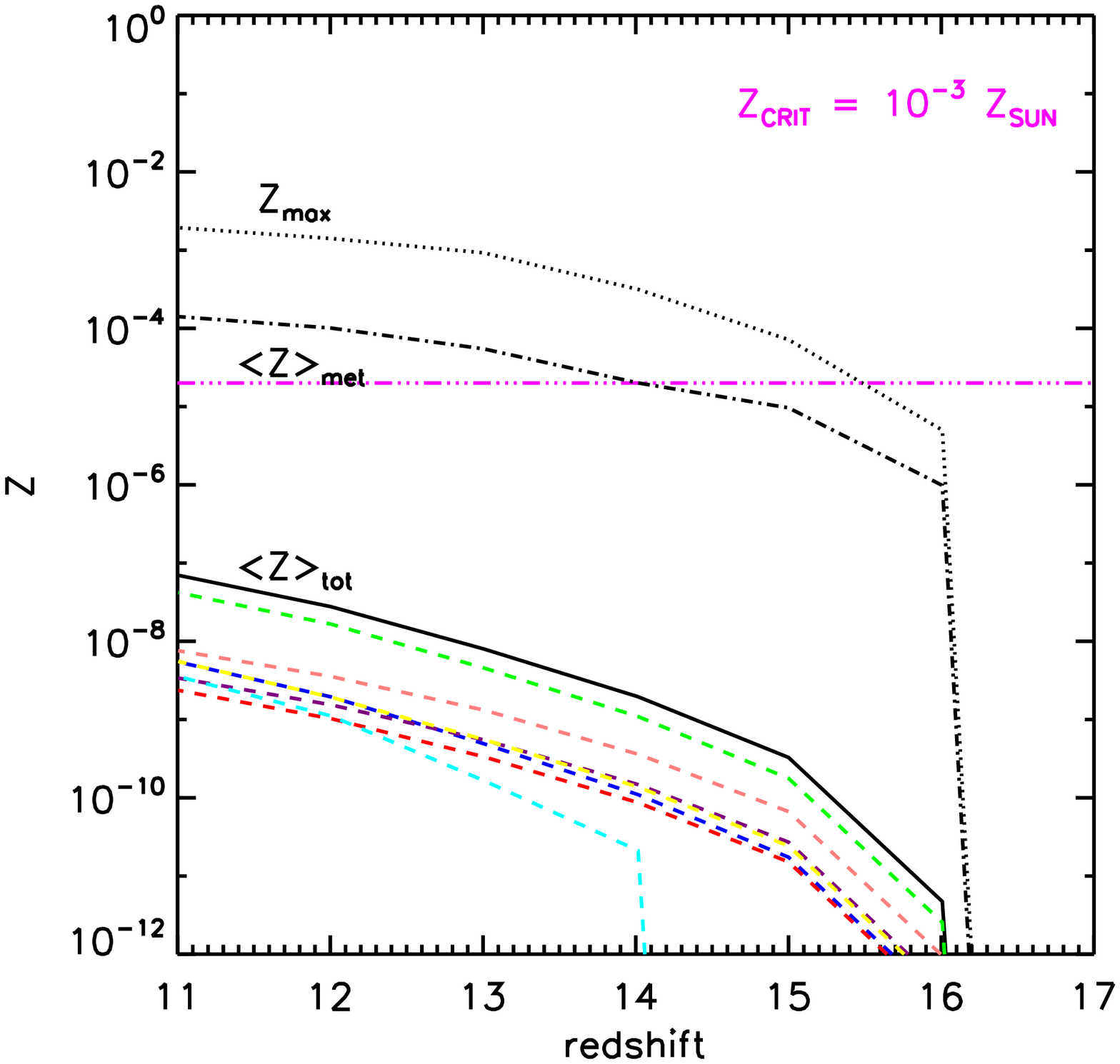}\hspace{-0.5cm}
\includegraphics[width=0.4\textwidth]{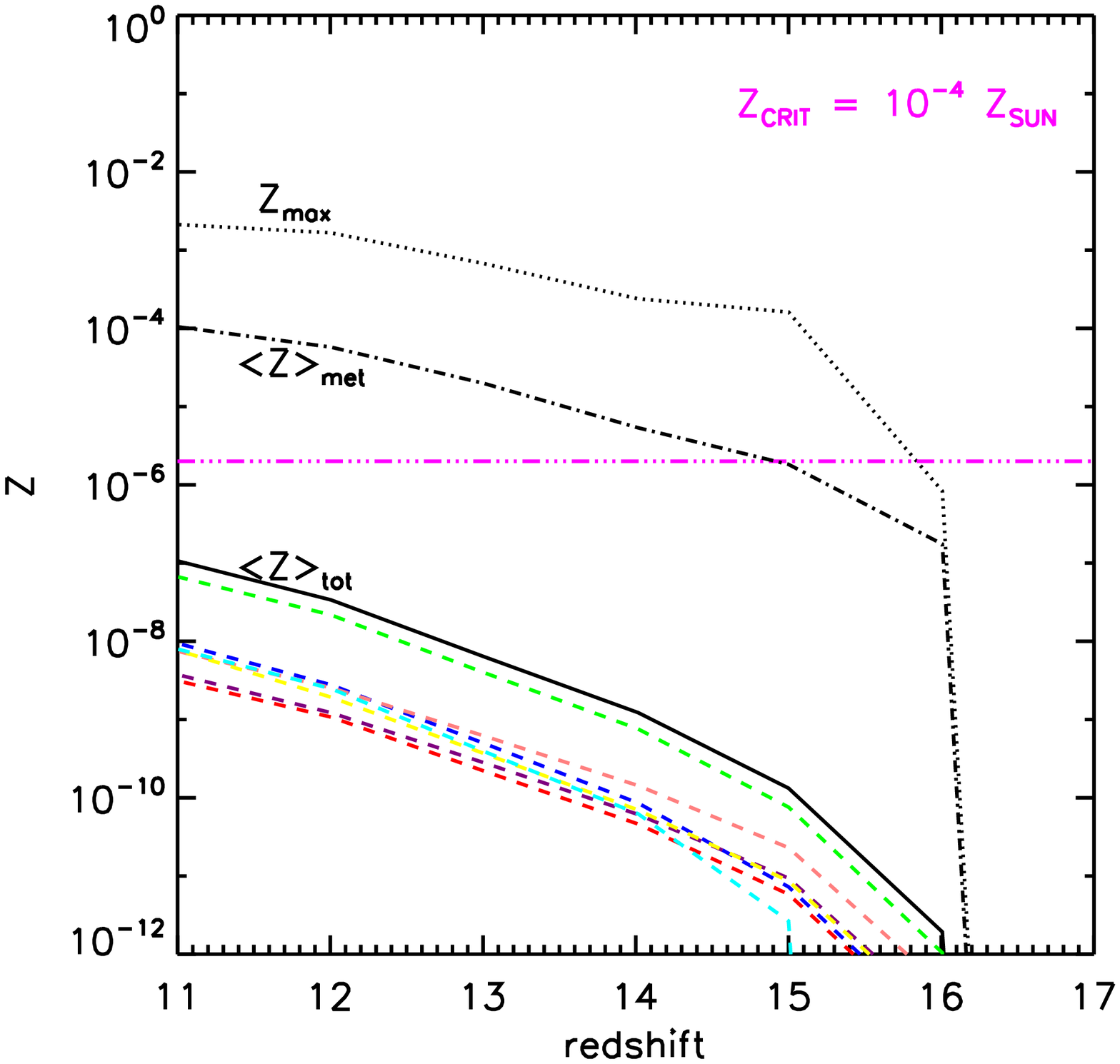}\\
\vspace{-0.5cm}
\includegraphics[width=0.4\textwidth]{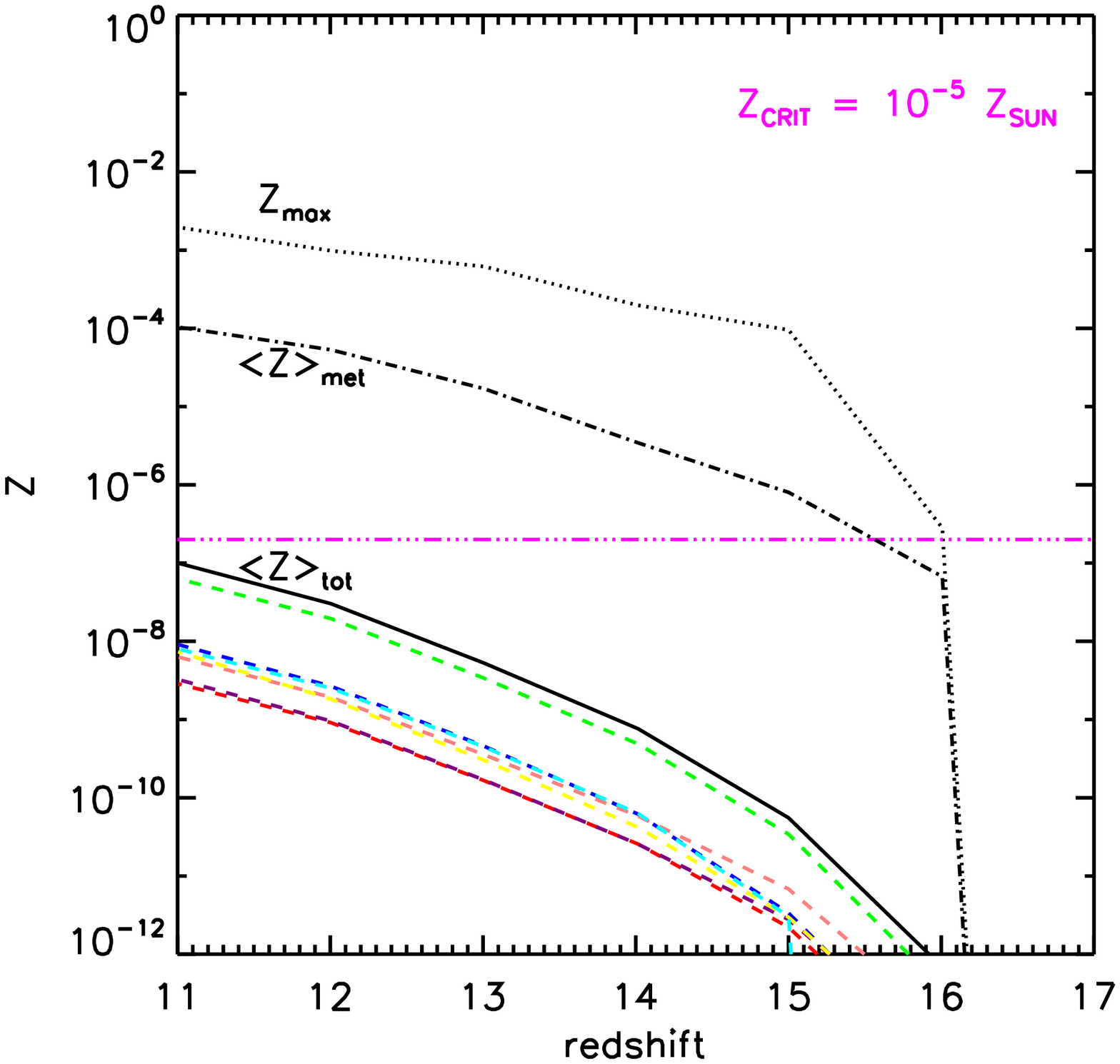}\hspace{-0.5cm}
\includegraphics[width=0.4\textwidth]{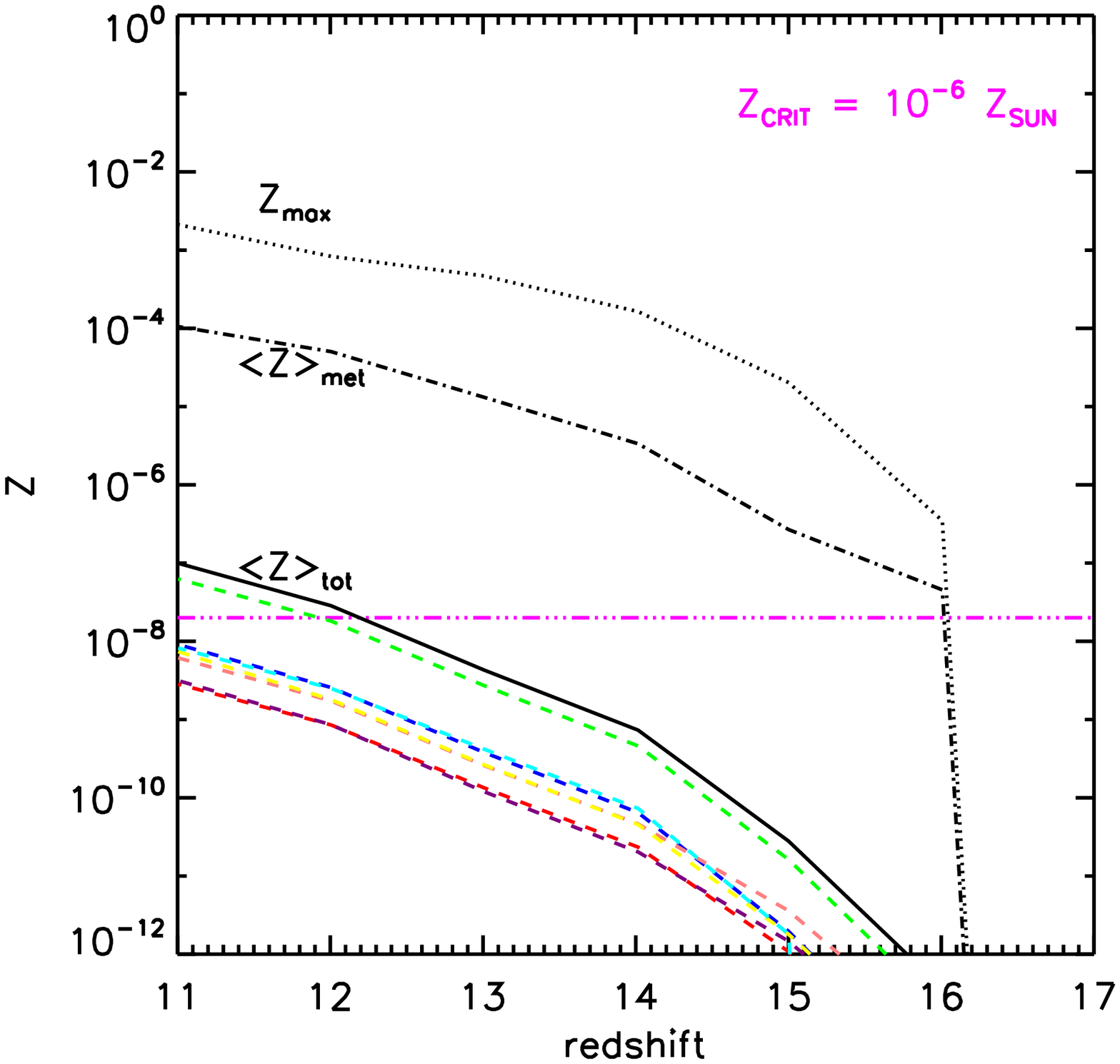}
\caption[Metallicity evolution]{\small Metal evolution as function of
  redshift for the $0.7\,{\rm Mpc}/h$ side box simulations, with
  $Z_{crit}=10^{-3}Z_\odot$ (upper-left panel), $10^{-4}Z_\odot$
  (upper-right panel), $10^{-5}Z_\odot$ (bottom-left panel), and
  $10^{-6}Z_\odot$ (bottom-right panel) case. The magenta horizontal
  dot-dot-dot-dashed line indicates, in each panel, the critical
  metallicity; the dotted line is the maximum metallicity; the
  dot-dashed line is the average metallicity of the spread metals; the
  solid line is the total metallicity averaged over the whole
  simulation box and the dashed lines the corresponding individual
  metallicities: oxygen (green), carbon (blue), magnesium (red),
  sulphur (purple), silicon (pink), iron (yellow), other metals (cyan)
  -- see also text in Sect. \ref{sect:sims}. }
\label{fig:Z_z}
\end{figure*}

From the previous results, it emerges that metal enrichment is a very 
patchy and inhomogeneous phenomenon. Never the less, as an indicative 
estimate of the global enrichment history of the Universe,
we present the evolution of the average metallicities found in the
simulations (Fig. \ref{fig:Z_z}). Following previous discussions though, we
expect large fluctuations\footnote{ Variances are comparable to the
  averages.} around the mean values reported here. In particular, for
each $Z_{crit}$ (labeled on the upper-right corner and shown by the
horizontal dot-dot-dot-dashed line in each panel), we show the maximum
metallicity reached at any redshift (dotted lines), the average
metallicity of the enriched regions (dot-dashed lines), the average
metallicity in the whole simulation box including primordial
unpolluted zones (solid line) and the corresponding individual
metallicities for the different metals tracked by the code (dashed
lines). 
\\ 
The first metals spread during the final stages of stellar
evolution leads to typical metallicities in the
  surrounding ISM of $Z\sim (10^{-5}-10^{-4})\,Z_\odot$, which,
shortly after, rise to values of $Z\sim (10^{-4}-10^{-3})\,Z_\odot$,
between redshift $z\sim 16$ and $z\sim 15$. Therefore, the critical
metallicity $Z_{crit}$ is easily overtaken, despite its precise,
actual determination. The leading element is always oxygen, as it is
the most abundant one produced by SNII and PISN explosions.
\\
For higher $Z_{crit}$ the popIII regime lasts longer, so that more massive
star explosions take place which can pollute the medium up to higher
metallicities, before popII-I star formation regime sets in. This
results in a quicker $Z$ increase in the very early stages (the
difference between the $Z_{crit} = 10^{-3}Z_\odot$ and $Z_{crit} =
10^{-6}Z_\odot$ cases is almost two orders of magnitude at $z\sim
15$). \\ Broadly speaking, a metallicity of $10^{-6}-10^{-5}\,Z_\odot$
is locally reached after only $\sim 10^5\,\rm yr$.
In $\sim 5\cdot 10^{6}\,\rm yr$,
it is possible to get $Z\sim 10^{-4}\,\rm Z_\odot$ or even $Z\sim
10^{-3}\,\rm Z_\odot$. An average pollution of $Z\sim Z_{crit}$ is
always obtained in a few $10^7\,\rm yr$ (i.e. $\Delta z\sim 1$, at
$z\sim 16$) after the onset of star formation (Fig. \ref{fig:Z_z}).
Once the critical metallicity is reached, popII-I star formation sets
in and contributes (on longer timescales) to metal enrichment, as
well.


\subsection{Star formation from population III and population II-I stellar generations}\label{sect:SFR}
\begin{figure}
\centering
\includegraphics[width=0.44\textwidth]{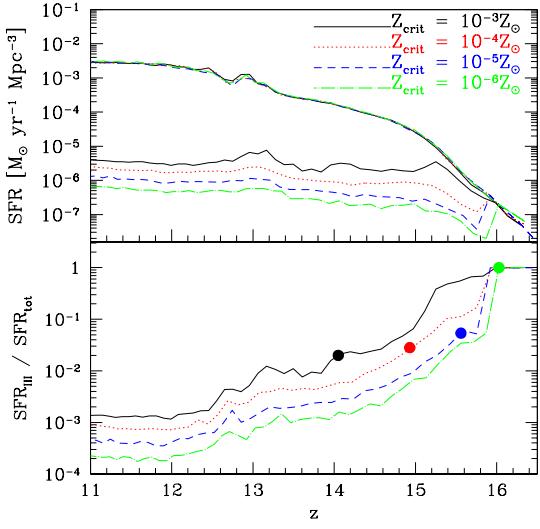}
\caption[Star formation]{\small {\it Upper panel}: redshift evolution
  of the average star formation rate densities (in $\rm M_\odot
  yr^{-1} Mpc^{-3}$) for the $0.7\,{\rm Mpc}/ {\it h}$ side box simulations.
  The four upper lines are the total star formation rate densities for
  $Z_{crit}=10^{-3}Z_\odot$ (solid line), $10^{-4}Z_\odot$ (dotted
  line), $10^{-5}Z_\odot$ (dashed line), and $10^{-6}Z_\odot$
  (dotted-dashed line). The four lower lines are the corresponding
  popIII contributions. {\it Bottom panel}: evolution of the ratio
  between the average popIII star formation rate density ($\rm
  SFR_{III}$) and the total star formation rate density ($\rm
  SFR_{tot}$), $\rm SFR_{III}/SFR_{tot}$. The bullets correspond to
  the redshift at which the average metallicity of the enriched
  particles equals the value of $Z_{crit}$ (as in Fig. \ref{fig:Z_z}).
}
\label{fig:sfr_ratio}
\end{figure}
In Fig. \ref{fig:sfr_ratio}, we plot the average\footnote{ The average
  is the arithmetic average of the star formation rates taken at each
  time-step and binned over 50 bins in redshift. } popIII star
formation rate, $\rm SFR_{III}$, and the average total star formation
rate, $\rm SFR_{tot}$. For all the simulations the onset of star
formation is at $z\simeq 16.3$, when the Universe is about $2.3\cdot
10^8\,\rm yr$ old (independently from $Z_{crit}$, which plays a role
only in the following epochs). As shown in Fig. \ref{fig:sfr_ratio},
at the very beginning popIII contribution is dominant, but rapidly
fades away and the popII-I regime is established.\\ For the
$Z_{crit}=10^{-3}Z_\odot$ case, the contribution is higher and it
decreases with $Z_{crit}$. In fact, the higher $Z_{crit}$, the higher
the number of star forming regions with $Z<Z_{crit}$. The total star
formation rate density is only mildly influenced by the exact value
adopted for $Z_{crit}$, meaning that popIII star formation does not
affect the global behaviour significantly and the bulk of star
formation is mainly led by popII-I stars. The reason for this is
simply understood in terms of timescales, as popIII stars have shorter
lifetimes (at most $\sim 10^6\,\rm yr$) and heavily pollute the medium
up to $Z_{crit}$ via PISN explosions. Therefore, after the first
bursts, it is much easier to match the conditions for standard popII-I
star formation, rather than metal free star formation. The period over
which star formation is dominated by popIII is slightly longer in the
$Z_{crit}=10^{-3}Z_\odot$ case and decreases gradually with
$Z_{crit}$. For $Z_{crit}=10^{-3}Z_\odot$, $\sim 100\%$ of the total
SFR is due to popIII stars in the redshift interval $z \sim 15-16$
(equivalent to a time interval of $\sim 2\cdot 10^7\,\rm yr$), while
the percentage decreases to an average value of roughly $0.1\%$ at
$z\sim 12$. In the $Z_{crit}=10^{-6}Z_\odot$ case, there is a sudden
drop below $\sim 10\%$ already at $z\sim 15-16$ and down to $\sim
0.01\%$ at $z\sim 12$. The intermediate cases are bracketed by the
former two regimes. These trends are consistent, because the time to
pollute the IGM to lower $Z_{crit}$ is shorter. The resulting
behaviour -- the smaller $Z_{crit}$, the earlier the transition from
popIII to popII-I dominated star formation -- is expected, because, as
mentioned, the time needed to pollute the IGM up to lower $Z_{crit}$
is shorter.\\ The dots on the lines of the average fractions of popIII
star formation rate densities, point to the redshifts when the average
enrichment reaches the level of $Z_{crit}$ (i.e. the abscissa of the
intersecting point between the dotted line and the horizontal line in
each panel of Fig. \ref{fig:Z_z}) for the different cases. At those
redshifts, the average contribution of popIII star formation has
already dropped of more than one order of magnitude. For the
$Z_{crit}=10^{-6}Z_\odot$ case, $Z_{crit}$ is reached in the polluted
regions when $\rm SFR_{III}/SFR_{tot}\sim 1$ and this ratio steeply
decreases afterwards. In the cases with higher $Z_{crit}$, the popII-I
regime develops simultaneously with the popIII regime, and the average
popIII contribution to the total star formation rate when
$Z\sim~Z_{crit}$ is below $\sim 10\%$.


\section{Parameter dependence}\label{sect:params}
In the following, we will investigate the effects on our results of
the main parameters (other than $Z_{crit}$) adopted in the simulations
and choose as reference the one with box dimension $0.7\,\rm Mpc/{\it
  h}$, $Z_{crit} = 10^{-4} Z_\odot$, $n_{\rm H, th} = 70 \,\rm
cm^{-3}$ and Salpeter slope for popIII stars in the range [$\rm 100
  M_\odot, 500 M_\odot$]. We have also checked the effects of
  adopting a different implementation of SPH \cite[the relative
    pressure SPH, {\it rp}SPH, according to][]{Abel2010}, which only
  considers pressure gradients in force calculations and performes
  closer to grid-based codes. We found no strong deviations in the
  general gas state \cite[as also expected from the discussion in
    Sect.~3.8 of][]{Abel2010} and even in the star formation rates.
  Metal pollution at early times is slightly more scattered, due to
  the capability of the {\it rp}SPH implementation of better resolving
  instabilities. Overall, there are no substantial changes in our results.

\subsection{Different population III IMF's and yields}\label{sect:imf}

\begin{figure}
\centering
\includegraphics[width=0.44\textwidth]{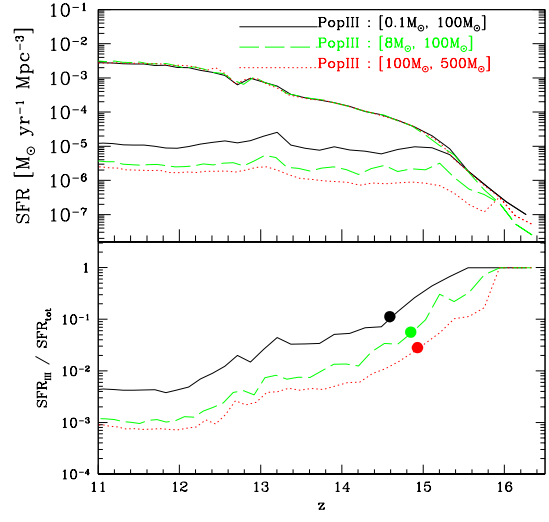}
\caption[Dependence of SFR on the threshold]{\small {\it Upper panel}:
  redshift evolution of the average star formation rate densities (in
  $\rm M_\odot yr^{-1} Mpc^{-3}$) for the $0.7\,\rm Mpc/ {\it h}$ side box
  simulations with $Z_{crit}=10^{-4} Z_\odot$. The upper lines are the
  total star formation rate densities for different popIII IMF mass
  range: $\rm [0.1~M_\odot, 100~M_\odot]$ (solid lines), $\rm
  [8~M_\odot, 100~M_\odot]$ (dashed lines), and $\rm [100~M_\odot,
    500~M_\odot]$ (dotted lines). The lower lines are the
  corresponding popIII contributions. {\it Bottom panel}: evolution of the ratio
  between the average popIII star formation rate density ($\rm
  SFR_{III}$) and the total star formation rate density ($\rm
  SFR_{tot}$), $\rm SFR_{III}/SFR_{tot}$. The bullets correspond to
  the redshift at which the average metallicity of the enriched
  particles equals the value of $Z_{crit}$.
}
\label{fig:imfrange}
\end{figure}
\begin{figure}
\centering
\includegraphics[width=0.44\textwidth]{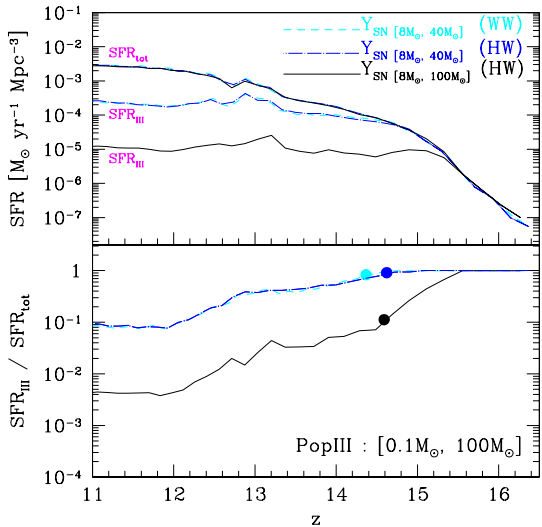}
\caption[Dependence of SFR on the threshold]{\small {\it Upper panel}:
  redshift evolution of the average star formation rate densities (in
  $\rm M_\odot yr^{-1} Mpc^{-3}$) for the $0.7\,\rm Mpc/ {\it h}$ side box
  simulations with $Z_{crit}=10^{-4} Z_\odot$. The upper lines are the
  total star formation rate densities for different popIII metal
  yields, and fixed the popIII IMF mass range at $\rm [0.1~M_\odot,
    100~M_\odot]$: primordial SN yields for stars of $\rm [8~M_\odot,
    40~M_\odot]$ \cite[case Z=0 in][]{WW1995} (dashed lines),
  primordial SN yields for stars of $\rm [8~M_\odot, 40~M_\odot]$
  \cite[][]{HegerWoosley2008} (dot-dashed lines), and primordial SN
  yields for stars of $\rm [8~M_\odot, 100~M_\odot]$
  \cite[][]{HegerWoosley2008} (solid lines). The lower lines are the
  corresponding popIII contributions. {\it Bottom panel}: evolution of
  the ratio between the average popIII star formation rate density
  ($\rm SFR_{III}$) and the total star formation rate density ($\rm
  SFR_{tot}$), $\rm SFR_{III}/SFR_{tot}$. The bullets correspond to
  the redshift at which the average metallicity of the enriched
  particles equals the value of $Z_{crit}$. }
\label{fig:imfyields}
\end{figure}
In order to account for the uncertainties on the popIII IMF, we
explore first the role of the slope of the popIII IMF, which so far
has been fixed to the value $s=-1.35$.
\\ 
There are no robust arguments
(neither theoretical nor observational) supporting the above
assumption and different slopes or shapes are not excluded. None the
less, by changing slope, the expected fraction of PISN is not
dramatically modified, as its value is always around $\sim 0.4$ for
any $s$ between roughly $-1$ and $-3$. Thus, the results on
metallicity evolution and star formation rates are not altered at
all.
\\ 
Differently, the chosen mass range and yields of popIII stars
could play a more important role, since they directly affect metal
pollution and star formation history. The existence of popIII stars
with masses below $\rm \sim~10^2~M_\odot$ is supported by numerical
simulations \cite[e.g.][]{Yoshida2006, Yoshida_et_al_2007,CampbellLattanzio2008,SudaFujimoto2010}, which find
that it is possible to fragment primordial clouds down to $\rm\sim 10~M_\odot$.  
According to \cite{WW1995}, $Z=0$ massive stars can die as SN if their mass is 
smaller than $\rm 40~M_\odot$; above this limit, they collapse into black holes. 
Recent updates, though, suggest that such stars can die as SN even if their masses 
are larger, up to $\rm 100~M_\odot$ \cite[][]{HegerWoosley2008}. We have investigated
the implications of the uncertainties on our results.
\\ 
In Fig. \ref{fig:imfrange}, together with the results for our reference run (dotted lines),
we show the trends for different popIII IMF mass ranges (while we keep the same slope),
assuming metal yields for primordial, massive SN from \cite{HegerWoosley2008}\footnote{
We use the tabulated post-supernova yields \cite[Table 8 in][]{HegerWoosley2008}
for the $\rm 10~M_\odot-100~M_\odot$ stars, explosion energy of 1.2~Bethe, 
standard mixing prescriptions, S=4 piston parameter (S.~Woosley, private communication). }.
We plot star formation rates and the corresponding popIII contribution for popIII
IMF mass ranges of: $\rm [0.1~M_\odot, 100~M_\odot]$ (solid lines),
$\rm [8~M_\odot, 100~M_\odot]$ (dashed lines), and $\rm [100~M_\odot, 500~M_\odot]$ 
(dotted lines). The latter case is our reference run
and its popIII contribution is always lower than the other ones. 
The reason is the extremely short lifetime of stars with masses in the
range $\rm [100~M_\odot, 500~M_\odot]$ (and in particular in the PISN
range), which pollute the surrounding medium in a time-lag much
shorter than the one needed to longer-living $\rm\sim 10~M_\odot$
stars. Therefore, metal pollution and the transition to popII-I regime
happen earlier, with a residual popIII contribution of a
factor of 3 or 4 smaller. The difference between the extreme case of stellar
masses in the range $\rm [0.1~M_\odot, 100~M_\odot]$ and the case of
stellar masses in the range $\rm [8~M_\odot, 100~M_\odot]$ is simply
due to the different normalization of the IMF, since yields and
massive-SN lifetimes are the same. Indeed, the SN fractions are $\sim
0.14$ and 1, respectively. Thus, as in the latter case
$100\%$ of the stars dies as SN, their enrichment results slightly
higher, and the popIII contribution drops faster than in the former
one (see also the bullets in the lower panel).
\\
Finally, we investigate the effects of different metal yields,
according to \cite{WW1995} and \cite{HegerWoosley2008}. As the
massive-SN mass range differs in the two works, in order to have a
fair comparison between the yields only and to exclude normalization
effects, we fix the popIII IMF range as $\rm [0.1~M_\odot,
  100~M_\odot]$ and the massive-SN mass range as $\rm
[8~M_\odot, 40~M_\odot]$. In Fig. \ref{fig:imfyields}, we compare the
results for the \cite{WW1995}'s yields (dashed lines) and for the
\cite{HegerWoosley2008}'s ones (dot-dashed lines).
For sake of completeness, we re-plot also the
results for the IMF mass range of $\rm [0.1~M_\odot, 100~M_\odot]$ and
the massive-SN range of $\rm [8~M_\odot, 100~M_\odot]$
\cite[][]{HegerWoosley2008}.
Due to the longer lifetimes of $\rm 10~M_\odot-40~M_\odot$ stars 
(dashed lines and dot-dashed lines), with respect to PISN and $\rm
100~M_\odot$ star lifetimes (solid lines), the popIII regime can now last longer
(compare bullets in Fig. \ref{fig:imfyields} with the ones in Fig.
\ref{fig:sfr_ratio} or in Fig. \ref{fig:imfrange}), and contribute up
to $10\%$ the total SFR at $z\sim 11$: some order of magnitude more
than all the other cases.
By directly comparing the effects of adopting different yields, in the 
same mass range, it comes out that their impact is negligible and there
are no significant changes in the star formation history.


\subsection{Density threshold for star formation}\label{sect:threshold}
\begin{figure}
\centering
\includegraphics[width=0.44\textwidth]{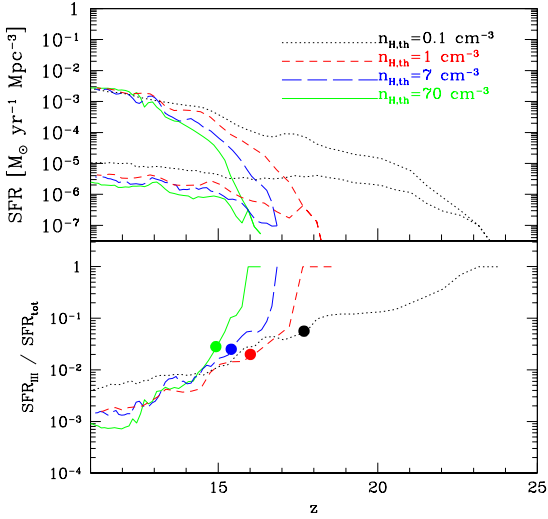}
\caption[Dependence of SFR on the threshold]{\small {\it Upper panel}:
  redshift evolution of the average star formation rate densities (in
  $\rm M_\odot yr^{-1} Mpc^{-3}$) for the $0.7\,\rm Mpc/h$ side box
  simulations with $Z_{crit}=10^{-4} Z_\odot$. The upper lines are the
  total star formation rate densities for different H-density
  thresholds for star formation, $n_{\rm H, th}= $ $0.1 \,\rm cm^{-3}$
  (dotted lines), $1 \,\rm cm^{-3}$ (short-dashed lines), $7 \,\rm
  cm^{-3}$ (long-dashed lines), and $70 \,\rm cm^{-3}$ (solid lines).
  The four lower lines are the corresponding popIII contributions.
  {\it Bottom panel}: evolution of the ratio between the average
  popIII star formation rate density ($\rm SFR_{III}$) and the total
  star formation rate density ($\rm SFR_{tot}$), $\rm
  SFR_{III}/SFR_{tot}$. The bullets correspond to the redshift at
  which the average metallicity of the enriched particles equals the
  value of $Z_{crit}$. }
\label{fig:threshold}
\end{figure}
We now address how the transition from popIII to popII-I depends on
the choice of the density threshold for star formation. Indeed, it is
usual to assume that star formation takes place only in converging
flows where a given number density is reached. Because different
thresholds lead to different onsets of star formation \cite[see
  also][]{Maio2009}, this could also affect the properties of the
transition. For this reason, we re-run the reference simulation
($n_{\rm H,th} = 70\,\rm cm^{-3}$) assuming density thresholds of
$n_{\rm H, th} = 7,1,0.1\,\rm cm^{-3}$.
Also in these cases the enrichment is very fast and the critical level in polluted
environments is always reached between redshifts $\sim 15$ and 18,
very close to the value of our fiducial case, $\sim 15$ (see Fig.
\ref{fig:Z_z}). Obviously, the higher the density threshold the later
the onset of star formation, as the gas needs more time to condense.
\\ 
In Fig. \ref{fig:threshold} we show the star formation rates and
the contribution from popIII stars for different density thresholds.
The asymptotic amplitude of the total SFR is not affected, but the
contribution from popIII stars has different slopes for different
$n_{\rm H, th}$: the higher the threshold, the steeper the evolution
of $\rm SFR_{III}/SFR_{tot}$, with this ratio decreasing of $\sim 3$
orders of magnitude in redshift intervals of $\Delta z\sim$
15, 8, 6, 4, for $n_{\rm H, tr} = $ 0.1, 1, 7, 70, respectively. 
\\ 
It is evident that the difference between the $70\,\rm cm^{-3}$, $7\,\rm
cm^{-3}$, and $1\,\rm cm^{-3}$ cases is quite small, with star
formation rates almost overlapping since the beginning of the process.
The reason is that such values fall in correspondence of the gas
cooling branch, which is extremely fast and therefore the effects on
the final results are very little. For the 0.1$\,\rm cm^{-3}$ case,
instead, we find discrepancies at higher redshift due to the
unresolved isothermal peak. This is due to the fact that (unphysical)
low-density threshold models convert gas into stars quite regularly
and smoothly, when it is still shock-heating in the hosting
dark-matter potential well. But, if high-density thresholds are used,
the dense gas is converted into stars later on, only after the
loitering, isothermal phase (during which it just gets accumulated at
temperatures around $\sim 10^4\,\rm K$) and the catastrophic run-away
cooling. So, large amounts of particles undergo star formation in a
shorter time, after which the SFR catches up and converges to the
low-density-threshold cases. In the meantime, the critical metallicity
is more quickly reached, as well. This also explains the differences
among the slopes of the decrement. The higher the density thresholds,
the higher the gas accumulated and catastrophically converted into
stars, the higher the enrichment, and the steeper the ratio $\rm
SFR_{III} / SFR_{tot}$. 
\\ 
The popIII/popII-I transition (see bullets)
happens in all the cases when the contribution of SFR$_{\rm III}$ has
dropped roughly between $10^{-2} - 10^{-1}$, and at redshift $z\sim
10$ it is $\sim 10^{-3}$. 
\\ 
These conclusions support the fact that
the transition from popIII to popII-I star formation regime is very
fast, independently from the detailed numerical parameters, as long as
the gas isothermal phase is resolved.


\subsection{Box size}\label{sect:largerboxes}

In this section, we discuss the effect on our results of different
choices for the box size. In the next, we will show metallicity
properties (Sect. \ref{MetallicityLB}) and star formation rate
densities (Sect. \ref{StarFormationLB}) for simulations with the
same parameters used for our reference simulation but with a box size
in comoving units of $5\,{\rm Mpc}/h$ and $10\,{\rm Mpc}/h$.
Properties and resolution are also summarized in Table \ref{tab:sims}.
Because of the lower resolution of such boxes (gas particle masses are
of the order $\sim 10^4 - 10^5 {\rm M_\odot}/h$), we cannot adopt a
very-high-density threshold (of $\sim 10^2\rm\,cm^{-3}$ or more),
therefore, to make a meaningful comparison we show results for
$n_{\rm H, th}=7\,\rm cm^{-3}$.


\subsubsection{Metallicity}\label{MetallicityLB}

\begin{figure}
\begin{center}
\includegraphics[width=0.44\textwidth]{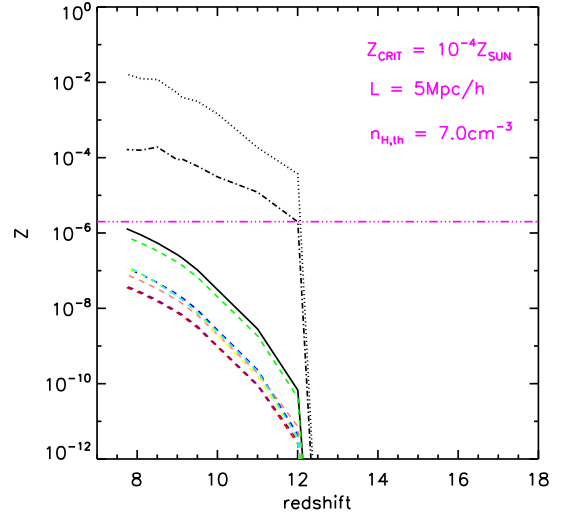}\hspace{-0.5cm}
\caption[Metallicity]{\small Average metallicities in the simulation
  of the box with a comoving side length of 5 Mpc/{\it h} and
  $Z_{crit}=10^{-4}Z_\odot$. The density threshold is $7\,\rm
  cm^{-3}$. The magenta horizontal dot-dot-dot-dashed line indicates
  the critical metallicity; the dotted line is the maximum
  metallicity; the dot-dashed line is the average metallicity of the
  spread metals; the solid line is the total metallicity averaged over
  the whole simulation box; and the dashed lines the corresponding
  individual metallicities (see also Fig. \ref{fig:Z_z}). }
\label{fig:Z_z_largerbox5}.
\end{center}
\end{figure}
\begin{figure}
\begin{center}
\includegraphics[width=0.44\textwidth]{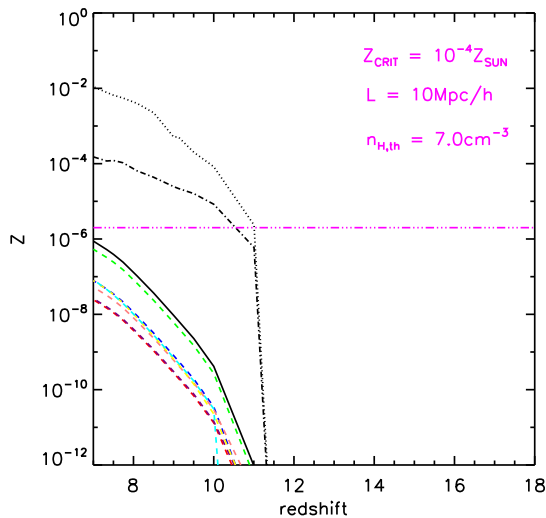}\hspace{-0.5cm}
\caption[Metallicity]{\small Average metallicities in the simulation
  of the box with a comoving side length of 10 Mpc/{\it h} and
  $Z_{crit}=10^{-4}Z_\odot$. The density threshold is $7\,\rm
  cm^{-3}$. The magenta horizontal dot-dot-dot-dashed line indicates
  the critical metallicity; the dotted line is the maximum
  metallicity; the dot-dashed line is the average metallicity of the
  spread metals; the solid line is the total metallicity averaged over
  the whole simulation box and the dashed lines the corresponding
  individual metallicities (see also Fig. \ref{fig:Z_z}).}
\label{fig:Z_z_largerbox10}.
\end{center}
\end{figure}
The global behaviour is patchy and similar to the maps already shown
for the $0.7~\rm Mpc/{\it h}$ cases (Figs.
\ref{fig:maps_metallicity_zcrit},
\ref{fig:maps_metallicity_comparison},
\ref{fig:maps_metallicity_comparison_2col}). The only difference is
the larger sampling of star forming regions with a wider statistical
significance. 
\\ 
In Figs. \ref{fig:Z_z_largerbox5} and
\ref{fig:Z_z_largerbox10}, we plot metallicity evolution for the
simulations with 5~Mpc/{\it h} and 10~Mpc/{\it h} box side,
respectively. In both cases, the critical metallicity is reached quite
quickly and metal-rich regions have $Z > Z_{crit}$ already by redshift
$\sim 10$. The rapid attainment of $Z_{crit}$ is independent of the
resolution, which instead influences the onset of star
formation. That is the reason why, in the 10~Mpc/{\it h} box side, star
formation is slightly delayed. As in Fig. \ref{fig:Z_z}, we also plot the
maximum (dotted lines) and average metallicity in the whole simulation
box (solid lines) with the different metal contributions (dashed
lines). We find that the first can easily reach solar values, because
of strong pollution due to the high PISN metal yields; the second has
$Z<Z_{crit}$ at any redshift above $\sim 7$. These fluctuations are 
reflected in the high patchiness of early metal spreading.
The metal pollution histories do not show large discrepancies and, in spite of 
differences in resolution of more than one order of magnitude for the SPH masses,
an overall agreement between larger (5 or 10 Mpc/$h$) and smaller (0.7
Mpc/$h$) boxes is reached. Thus, the general results are
independent from the particular cosmological sampling.


\subsubsection{Star formation from population III and population II-I stellar generations}\label{StarFormationLB}
\begin{figure}
\centering
\includegraphics[width=0.44\textwidth]{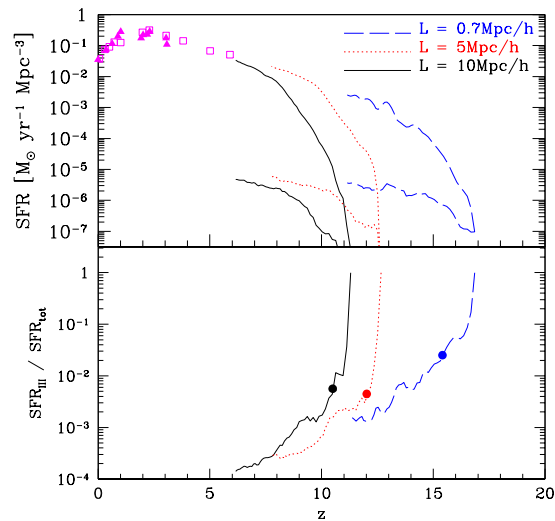}
\caption[Cosmological star formation rate densities]{\small {\it Upper
    panel}: star formation rate densities for the simulations with a 
    comoving box side of $0.7\,\rm Mpc/{\it h}$ (dashed lines), $5\,{\rm
    Mpc/}h$ (dotted lines) and $10 {\rm Mpc/}h$ (solid lines). The
  density threshold is $7 \,\rm cm^{-3}$. For each of them, we show
  the total star formation rate density (upper lines) and the popIII
  one (lower lines). Data points \cite[][]{Reddy_et_al_2008} are
  different determinations of the star formation rate density: square
  symbols are the UV determinations, triangular symbols are the IR
  determinations. {\it Bottom panel}: evolution of the ratio between
  the average popIII star formation rate density ($\rm SFR_{III}$) and
  the total star formation rate density ($\rm SFR_{tot}$), $\rm
  SFR_{III}/SFR_{tot}$. The bullets correspond to the redshift at
  which the average metallicity of the enriched particles equals the
  value of $Z_{crit}$. }
\label{fig:sfr_resolution_ratio}
\end{figure}
We will now discuss the behaviour of the star formation rates for the
simulations previously introduced. 
\\ 
The average star formation rate
densities are shown in the upper panel of
Fig.~\ref{fig:sfr_resolution_ratio}, for the boxes of $0.7\,{\rm
  Mpc}/h$ (dashed lines), $5\,{\rm Mpc}/h$ (dotted lines), and
$10\,{\rm Mpc}/h$ (solid lines). The upper lines refer to the
total star formation rate densities and the ones below to the
corresponding popIII star formation rate densities \cite[data points
  are by ][]{Reddy_et_al_2008}. Overall, they seem to connect fairly
well with the observed trend at lower redshift.
\\ 
In the bottom panel, we present the
ratios of popIII versus total star formation rate. Independently from
the box size, the popIII contribution is unity at the very beginning
(i.e. $\sim 100\%$ of the SFR is due to popIII) and, for all the
cases, rapidly decreases below $\sim 1\%$ at redshift $z\sim 11$, and
below $\sim 0.1\%$ at redshift $z \lesssim 11$: a time-lag much
shorter than a Gyr. This means, consistently with what said in Sect.
\ref{sect:results}, that the bulk of popIII episodes happpens in a
short time after the onset of star formation and it dominates the
cosmological SFR for a limited period at high redshift. At lower
redshift, only isolated regions could still have pristine environment
and host popIII star formation (which is none the less several order
of magnitude smaller that the total SFR).
\\
A general feature of the large box simulations is
the slight delay of the onset of star formation: this is due to the
limited capabilities of the bigger SPH particles to condense, cool,
and collapse. Indeed, molecular cooling is not efficient when the SPH
masses are so high ($\sim 10^4-10^5 {\rm M_\odot}/h$) and the
primordial gas in-fall is not well resolved. The qualitative agreement
of these results is encouraging, though.


\section{Discussion and conclusions}\label{sect:disc}
In this work, we perform numerical simulations of early structure
formation, including both primordial, molecular evolution and metal
enrichment from stellar death, to study the transition from an early, massive
star formation mode to a more standard one, regulated by the gas metallicity.
We follow \cite[see details in][]{Maio2007,TBDM2007} the abundances of 
e$^-$, H, H$^+$, H$^-$, He, He$^+$, He$^{++}$, H$_2$, H$_2^+$, D, D$^+$, HD, HeH$^+$, C, O, Mg, S, Si, Fe,
and use different initial stellar-mass functions, metal yields, and critical metallicities for the transition from a popIII to a popII star formation mode, $Z_{crit}$.
At the present, our work is
the only one dealing with detailed chemical evolution, from early
molecule creation to the later stages of star formation, and, at the
same time, allowing to trace simultaneous, different stellar populations, 
according to the underlying metallicity.
\\
Indeed, early structure formation can be accurately modeled only with a proper
treatment of both the chemistry of primordial molecules and the metal
enrichment. In fact, the main influence of chemical evolution on the
following generations of structures is via metal pollution (chemical
feedback). This event can completely alter the cooling properties of
the gas and thus the modalities of star formation, inducing a
transition from a top-heavy to a standard IMF. The transition is
believed to happen when the gas is enriched above $Z_{crit}$, which
allows fragmentation below the typical scales determined by primordial
molecular cooling. Because of our ignorance of the features of early
dust formation and the lack of precision of many atomic and molecular
data, the exact determination of $Z_{crit}$ is still elusive;
reasonable values should range between $10^{-6}\,Z_\odot$ and
$10^{-3}\,Z_\odot$ \cite[][]{Bromm_et_al_2001,Schneider_et_al_2002}.
\\ 
In our simulations we assume four different values of $Z_{crit}$ in
the above range: $10^{-6}, 10^{-5}, 10^{-4}$, and $10^{-3}\,Z_\odot$.
From our investigation, some common features emerge. In general, metal
pollution, independently from the parameters adopted in the
simulations, is very patchy, with excursions of orders of magnitudes
at all redshifts. This is consistent with any simulation including
metal enrichment
\cite[][]{Raiteri1996,Gnedin1998,Mosconi2001,Lia2002a,Lia2002b,KawataGibson2003,Kobayashi2004,RicottiOstriker2004,Cecilia_et_al_2005,TBDM2007,Oppenheimer_et_al_2010}.
In addition, the gas is easily enriched above $Z_{crit}$. For this
reason, the average contribution from pristine, metal-free (or $Z <
Z_{crit}$) stars to the total cosmic star formation density is
dominant only in the very early phases of structure formation, while
it drops below $\sim 10^{-3}$ quite rapidly, after the explosion of
the first pair-instability supernov{\ae} and their metal ejection. In
fact, PISN explosions which follow the death of the first, metal-free
or very-metal-poor stars, are the main responsibles for enriching the
surrounding medium up to a minimum level of $\sim 10^{-4}\,Z_\odot$.
This means that nearby star forming regions have a very high
probability of being polluted above $Z_{crit}$ (as seen also in Fig.
\ref{fig:Z_z}), while popIII star formation can still occur farther
away, in rare, isolated regions with pristine or low-metallicity gas
(e.g. Fig. \ref{fig:maps_metallicity_comparison_2col}). None the less,
this would not be the dominant star formation regime. 
\\
Our findings hold regardless of the value of $Z_{crit}$, the popIII IMF
adopted, and the numerical parameters involved in modeling star
formation (as shown in Sect. \ref{sect:imf}, \ref{sect:threshold},
and \ref{sect:largerboxes}).
Differences are found if different IMF mass ranges for primordial stars are used.
Because of the short life and the high metal yields of early, massive SN,
the popIII regime contributes, in any case, only slightly to the global SFR 
\cite[as expected by e.g. ][]{RicottiOstriker2004}, since the early pollution
events quickly raise $Z$ above $Z_{crit}$, independently from the detailed
prescriptions.
\\
The simulations were performed using a standard $\Lambda$CDM
cosmology, but slightly different parameters would not change the
general picture. We have checked this, by running simulations with
$\Omega_{0m}=0.26$, $\Omega_{0\Lambda}=0.74$, $\Omega_{0b}=0.0441$,
$h=0.72$, $\sigma_8=0.796$, and $n=0.96$ \cite[WMAP5 data,
][]{wmap5_2008} and found that the same results hold, albeit shifted
by a $\Delta z\sim$ of a few to lower redshift (because of the smaller
$\Omega_{0m}$).
\\
One last comment about the role of dust production from SNII and/or PISN:
one of the main uncertainties in determining $Z_{crit}$.
According to our findings, given the strength and rapidity of metal enrichment
(see e.g. Figs. \ref{fig:Z_z}, \ref{fig:Z_z_largerbox5} and
\ref{fig:Z_z_largerbox10}), details about dust and its impact on
$Z_{crit}$ are not relevant for the general process and for our understanding 
of the transition from primordial, popIII regime to present-day-like, popII-I regime 
(even if they could be of some interest on very local scales).
The entire process, in fact, is dominated by metal pollution and the strong yields
of early, massive stars.
\\
We have to stress some {\it caveats}, though.
Ejection of particles into the IGM is an unknown process.
We have assumed winds originated from stars (kinetic feedback), 
but different mechanisms (like gas stripping, shocks, 
thermal heating from stellar radiation, etc.) could play a role, as well, 
mostly at high redshifts, when objects are small and can easily loose part 
of their baryonic content.
In addition, diffusion and conduction will probably alter the
smoothness of metal and molecule distribution, but these phenomena
have not been extensively studied, yet, and probably will depend on
many parameters: e.g., the way metal or gas particles are ejected and
mixed, how they are transferred away from the production sites, how
strong is the efficiency of such processes, just to mention a few. 
Some attempts to address such issues have been done
\cite[][]{Spitzer1962,CowieMcKee1977,Brookshaw1985,Sarazin1988,Monaghan1992,ClearyMonaghan1999,KlessenLin2003,Jubelgas_et_al_2004,Monaghan_et_al_2005,Wadsley_et_al_2008,Greif2009},
but much more realistic and detailed analyses are still strongly
needed. 
\\ 
To conclude, we have followed the structure formation
process from very early times to first star formation and subsequent
metal pollution. In the 1~Mpc side box simulations we have seen
that, after $\sim 2\times 10^8\,\rm yr$, molecular evolution leads the
very first bursts of star formation (popIII), but metal enrichment is
extremely fast (Figs. \ref{fig:maps_metallicity_zcrit},
\ref{fig:maps_metallicity_comparison}, and 
\ref{fig:maps_metallicity_comparison_2col}) in inducing
the nowadays observed star formation mode (popII-I). In fact, we
observe a steep increase of $Z$, with local values rapidly reaching
and overtaking $Z_{crit}$ (Fig. \ref{fig:Z_z}). Metal pollution
proceeds from the densest cores of star formation outwards, because of
supernova ejections from high-density to lower-density environments
(Figs. \ref{fig:maps_metallicity_comparison}, and
\ref{fig:maps_metallicity_comparison_2col}). 
Rare, unpolluted regions can still survive, determining the
simultaneous presence of the two star formation regimes, and $Z_{crit}$ can affect
the level of residual popIII star formation.
As a result of this rapid
pollution, we find that the average contribution of the popIII
component to the total star formation rate density is of a few times
$\sim 10^{-4}-10^{-3}$ (with a maximum of $\sim 10^{-2}-10^{-1}$) at
$z \sim 11$ (Figs. \ref{fig:sfr_ratio},\ref{fig:imfrange}, and 
\ref{fig:imfyields}).
This general picture is preserved
regardless of the precise value of the metallicity threshold $Z_{crit}$
the slope of the popIII IMF, and their yields, but is quite
sensitive to the popIII mass range. The change of the
critical density threshold for star formation influences little these
conclusions, but can affect the onset and the decrement of the popIII
contribution to the total SFR (e.g. Fig. \ref{fig:threshold}). We have
found similar results in larger-box simulations, either in the metal
enrichment features (Figs. \ref{fig:Z_z_largerbox5} and
\ref{fig:Z_z_largerbox10}) and in the star formation behaviour (Fig.
\ref{fig:sfr_resolution_ratio}).


\section*{acknowledgments}
U.~M. wishes to thank useful discussions with Claudio Dalla Vecchia, Massimo Dotti, 
Jarrett Johnson, Raffaella Schneider, and Stan Woosley.
He also thanks Naoki Yoshida and the travel support from the Grant-in-Aid for
Scientific Research (S) 20674003 by Japan Society for the Promotion of
Science. The simulations were performed by using the machines of the
Max Planck Society computing center, Garching
(Rechenzentrum-Garching). For the bibliografic research we have made
use of the tools offered by the NASA ADS, and by the JSTOR Archive.\\
More detailed maps and information can be found at the URL \href{http://www.mpe.mpg.de/~umaio/maps.html}{http://www.mpe.mpg.de/~umaio/maps.html}.


\bibliographystyle{mn2e}
\bibliography{bibl.bib}

\label{lastpage}
\end{document}